\documentclass[prd,amsmath,amssymb,showpacs,superscriptaddress,nofootinbib,twocolumn]{revtex4-1}
\usepackage{graphicx}
\usepackage{epsfig,graphics,subfigure,psfrag,amsmath,amssymb}
\usepackage{lineno}
\usepackage{dcolumn}
\usepackage{bm}
\usepackage{overpic}
\usepackage{xspace}
\usepackage{xcolor}
\usepackage{rotating}
\usepackage{color}
\usepackage{multirow}
\usepackage{colortbl}
\usepackage{epstopdf}
\usepackage{float}
\usepackage{makecell}
\usepackage[colorlinks,linkcolor=blue,anchorcolor=blue,citecolor=blue]{hyperref}
\newcommand{\gev}{\ensuremath{\mathrm{\,Ge\kern -0.1em V}}\xspace}
\newcommand{\mev}{\ensuremath{\mathrm{\,Me\kern -0.1em V}}\xspace}
\newcommand{\mevcc}{\ensuremath{{\mathrm{\,Me\kern -0.1em V\!/}c^2}}\xspace}

\def\fz#1       {\ensuremath{f_0({#1})}\xspace}

\usepackage{lineno}
\usepackage{hyperref}
\hypersetup{colorlinks = true,
            linkcolor = black,
            urlcolor = blue,
            citecolor = blue}

\begin{document}

\title{\bf \boldmath Search for the double Dalitz decays $\eta/\eta'\to e^+e^-\mu^+\mu^-$ and  $\eta^\prime \to \mu^+\mu^-\mu^+\mu^-$}

\author{
\begin{center}
M.~Ablikim$^{1}$, M.~N.~Achasov$^{4,c}$, P.~Adlarson$^{76}$, O.~Afedulidis$^{3}$, X.~C.~Ai$^{81}$, R.~Aliberti$^{35}$, A.~Amoroso$^{75A,75C}$, Y.~Bai$^{57}$, O.~Bakina$^{36}$, I.~Balossino$^{29A}$
% ...
\\
\vspace{0.2cm}
(BESIII Collaboration)\\
\vspace{0.2cm} {\it
$^{1}$ Institute of High Energy Physics, Beijing 100049, People's Republic of China\\
% ...
$^{3}$ Bochum  Ruhr-University, D-44780 Bochum, Germany\\
$^{4}$ Budker Institute of Nuclear Physics SB RAS (BINP), Novosibirsk 630090, Russia\\
% ...
$^{29}$ INFN Sezione di Ferrara, (A)INFN Sezione di Ferrara, I-44122, Ferrara, Italy; (B)University of Ferrara,  I-44122, Ferrara, Italy\\
% ...
$^{35}$ Johannes Gutenberg University of Mainz, Johann-Joachim-Becher-Weg 45, D-55099 Mainz, Germany\\
$^{36}$ Joint Institute for Nuclear Research, 141980 Dubna, Moscow region, Russia\\
% ...
$^{57}$ Southeast University, Nanjing 211100, People's Republic of China\\
% ...
$^{75}$ University of Turin and INFN, (A)University of Turin, I-10125, Turin, Italy; (B)University of Eastern Piedmont, I-15121, Alessandria, Italy; (C)INFN, I-10125, Turin, Italy\\
$^{76}$ Uppsala University, Box 516, SE-75120 Uppsala, Sweden\\
% ...
$^{81}$ Zhengzhou University, Zhengzhou 450001, People's Republic of China\\

\vspace{0.2cm}
% ...
$^{c}$ Also at the Novosibirsk State University, Novosibirsk, 630090, Russia\\
% ...

}

\end{center}
\vspace{0.4cm}
}

\begin{abstract}
\vspace{0.2cm}
Using a data sample of $(10087 \pm 44) \times {10^{6}}$ $J/{\psi}$ events collected with the BESIII detector, we search for the decays $\eta/\eta'\to e^+e^-\mu^+\mu^-$ and  $\eta^\prime \to \mu^+\mu^-\mu^+\mu^-$ via the radiative decays $J/{\psi}\to\gamma\eta$/$\gamma\eta'$. No excess of events over expected background is observed for any of the decays of interest. At 90\% confidence level, we report the first upper limits on the branching fractions of $\eta' \to e^{+}e^{-}\mu^{+}\mu^{-}$ and $\eta' \to \mu^{+}\mu^{-}\mu^{+}\mu^{-}$ to be $ 1.75 \times {10^{-6}}$ and $5.28 \times {10^{-7}}$, respectively. In addition, we set an upper limit on the branching fraction of $\eta \to e^{+}e^{-}\mu^{+}\mu^{-}$ to be $6.88 \times {10^{-6}}$, which improves the previous result by about two orders of magnitude.
\end{abstract}


\maketitle


\section{INTRODUCTION}

The second order electromagnetic decays $\eta/\eta^\prime\rightarrow \ell^+\ell^-
\ell^+\ell^-$ ($\ell=e,\mu$) proceed through an intermediate state of two virtual photons with each virtual photon undergoing internal conversion to a leptonic pair. Deviations between the measured rate and the rate predicted by quantum electrodynamics~(QED) are usually described in terms of a timelike transition form factor~(TFF), which is an important probe into the meson’s structure. In addition, the TFFs of the $\eta$ and $\eta^\prime$ play an important role in the evaluation of the hadronic light-by-light contribution to the muon anomalous magnetic moment~\cite{muon}.

Initial QED predictions~\cite{Jarlskog:1967fpu} of the double Dalitz decays of the $\eta$ were initially formulated about six decades ago, and assumed the unity of the TFF. Afterwards, many theoretical models, including the hidden gauge model~\cite{HG1,HG2,HG3}, the modified vector meson dominance model (VMD)~\cite{m_VMD1,m_VMD2,m_VMD3,VMD} and a data driven approach~\cite{DDA}, have been presented to describe the double Dalitz decays $\eta/\eta^\prime\rightarrow \ell^+\ell^-\ell^+\ell^-$. The corresponding predictions are summarized in Table~\ref{list0}. Experimentally, KLOE reported the first observation of $\eta\rightarrow e^+e^-e^+e^-$~\cite{KLOE:2011qwm}. Most recently, BESIII and CMS observed the decays $\eta^\prime\rightarrow e^+e^-e^+e^-$~\cite{BESIII:2022cul} and $\eta\rightarrow \mu^+\mu^-\mu^+\mu^-$~\cite{CMS:2023thf}, respectively, while studies of other $\eta/\eta^\prime\rightarrow \ell^+\ell^-\ell^+\ell^-$ decays are still scarce~\cite{PDG}. Only an upper limit on the branching fraction is available for $\eta \to e^{+}e^{-}\mu^{+}\mu^{-}$~\cite{KLOE:2008} and no experimental searches for $\eta^\prime \to \ell^{+}\ell^{-}\mu^{+}\mu^{-}$ have been performed so far. 

\begin{table*}[t]
\begin{center}
\caption{Theoretical predictions and experimental results on the 
branching fractions of $\eta/\eta^\prime\rightarrow \ell^+\ell^-\ell^+\ell^-$.}
\label{list0}
\begin{tabular}{ c c c c c }
     \hline
     \hline
    Decay&   Hidden gauge~\cite{VMD}&   Modified VMD~\cite{VMD}&   Data driven approach~\cite{DDA}&   Experimental result   \\
     \hline
${\eta\to e^{+}e^{-}e^{+}e^{-}}$ & {$2.680(13)\times 10^{-5}$} & {$2.668(13)\times 10^{-5}$} & {$2.17(2)\times 10^{-5}$} & {$2.40(22)\times 10^{-5}$~\cite{KLOE:2011qwm}}\\
${\eta' \to e^{+}e^{-}e^{+}e^{-}}$ & {$2.384(4)\times 10^{-6}$} & {$2.317(4)\times 10^{-6}$} & {$2.10(45)\times 10^{-6}$} & {$4.5(1)\times 10^{-6}$~\cite{BESIII:2022cul}}\\
${\eta\to\mu^{+}\mu^{-}\mu^{+}\mu^{-}}$ & {$3.992(27)\times 10^{-9}$} & {$3.797(26)\times 10^{-9}$} & {$3.98(15)\times 10^{-9}$} & {$5.0(8)\times 10^{-9}$~\cite{CMS:2023thf}}\\
${\eta' \to\mu^{+}\mu^{-}\mu^{+}\mu^{-}}$ & {$2.360(12)\times 10^{-8}$} & {$2.185(10)\times 10^{-8}$} & {$1.69(36)\times 10^{-8}$} & -\\
${\eta\to e^{+}e^{-}\mu^{+}\mu^{-}}$  & {$2.213(26)\times 10^{-6}$} & {$2.154(22)\times 10^{-6}$} & {$2.39(7)\times 10^{-6}$} & {$<1.6\times 10^{-4}$~\cite{KLOE:2008}}\\
${\eta' \to e^{+}e^{-}\mu^{+}\mu^{-}}$  & {$8.626(33)\times 10^{-7}$} & {$7.968(31)\times 10^{-7}$} & {$6.39(91)\times 10^{-7}$} & - \\
    \hline
    \hline
\end{tabular}
\end{center}
\end{table*}

As demonstrated in Ref.~\cite{Fang:2021hyq}, the BESIII experiment offers a unique opportunity to study the light mesons produced in $J/\psi$ decays by taking advantage of the largest $J/\psi$ data sample in the world. In this paper, we performed a search for the double Dalitz decays $\eta/\eta'\to e^+e^-\mu^+\mu^-$ and  $\eta^\prime \to \mu^+\mu^-\mu^+\mu^-$ using $(10087 \pm 44) \times {10^{6}}$ $J/\psi$ events~\cite{EVENTS} collected by the BESIII detector.

{
\section{BESIII DETECTOR AND MONTE CARLO SIMULATION}
\label{sec:BES}

The BESIII detector~\cite{ref::BesIII} records symmetric $e^+e^-$ collisions provided by the BEPCII accelerator~\cite{ref::collider} in the center-of-mass energy ($\sqrt{s}$) range from 2.0 to 4.95~GeV, with a peak luminosity of $1.1 \times 10^{33}\;\text{cm}^{-2}\text{s}^{-1}$ achieved at $\sqrt{s} = 3.773\;\text{GeV}$. The cylindrical core of the BESIII detector covers 93\% of the full solid angle and consists of a helium-based multilayer drift chamber~(MDC), a plastic scintillator time-of-flight system (TOF), and a CsI(Tl) electromagnetic calorimeter (EMC), which are all enclosed in a superconducting solenoidal magnet providing a 1.0~T magnetic field. The solenoid is supported by an octagonal flux-return yoke with resistive plate counter muon identifier modules interleaved with steel. The charged-particle momentum resolution at $1~{\rm GeV}/c$ is $0.5\%$, and the specific ionization energy loss d$E$/d$x$ resolution is $6\%$ for the electrons from Bhabha scattering. The EMC measures photon energies with a resolution of $2.5\%$ ($5\%$) at $1$~GeV in the barrel (end-cap) region. The time resolution of the TOF barrel part is 68~ps, while that of the end-cap part is 110~ps. The end-cap TOF system was upgraded in 2015 using multi-gap resistive plate chamber technology, providing a time resolution of 60~ps~\cite{Tof1,Tof2,Tof3}.
\setlength{\parskip}{0.18em}

Simulated samples of events are produced with a {\sc geant4}-based~\cite{Geant4} Monte Carlo~(MC) package, which includes the geometric description of the BESIII detector and the detector response.  These samples are used to determine the detection efficiency and to estimate the backgrounds. 
The simulation includes the beam energy spread and initial state radiation (ISR) in the $e^+e^-$ annihilations modeled with the generator {\sc kkmc}~\cite{Jadach01}, while the decays are simulated using {\sc evtgen}~\cite{EVT}.
Possible hadronic backgrounds are studied using a sample of $J/{\psi}$ inclusive events in which the known decays of the $J/{\psi}$ are modeled with branching fractions set to be the world average values from the particle data group~(PDG)~\cite{PDG}, while the unknown decays are generated with the LUNDCHARM model~\cite{LUN}. In order to describe the data well, some dedicated generators are developed based on the theoretical amplitudes for this analysis, such as $\eta / \eta' \to l^+l^-\mu^+\mu^-$~\cite{DIY1}, $\eta / \eta' \to \pi^+\pi^-l^+l^-$~\cite{DIY1}, $\eta / \eta' \to \gamma \l^+\l^-$~\cite{DIY2}, $\eta / \eta' \to \gamma \pi^+\pi^-$~\cite{DIY3}, which are based on the VMD model~\cite{VMD}. 
}

\section{EVENTS SELECTION}
\label{sec:selection}

The final states of interest are  $\gamma e^{+}e^{-}\mu^{+}\mu^{-}$ and $\gamma \mu^{+}\mu^{-}\mu^{+}\mu^{-}$.
Each event is required to contain four charged track candidates with zero net charge, and at least one photon candidate. Charged tracks detected by the MDC are required to be within a polar angle range of $|\rm{cos\theta}|\leq0.93$, where $\theta$ is defined with respect to the $z$-axis, which is the symmetry axis of the MDC. Each charged track is required to have the point of closest approach to the interaction point (IP) within $\pm{1}$ cm in the plane perpendicular to the beam direction and within $\pm{10}$ cm in the beam direction.

\setlength{\parskip}{0.18em}
Photons are reconstructed from showers in the EMC exceeding a deposited energy of $25$ MeV in the barrel region ($|\rm{cos\theta}|<0.8$) and $50$ MeV in the endcap regions ($0.86<|\rm{cos\theta}|<0.92$). The opening angle between the position of the shower and the charged tracks extrapolated to the EMC must be greater than $15$ degrees. Finally, the photon candidates are required to arrive in the EMC within 700 ns from the event start time in order to reduce the backgrounds unrelated to the event. 

To identify the charged tracks and select the best photon when additional photons are found in an event, particle identification (PID) requirements are applied using the TOF and d$E$/d$x$ information. Furthermore, a four-constraint (4C) kinematic fit is performed by imposing energy and momentum conservation under the hypothesis of $\gamma e^+e^-\mu^+\mu^-$ and $\gamma \mu^+\mu^-\mu^+\mu^-$, respectively. The combination with the minimum value of $\chi^2_{l^+l^-\mu^+\mu^-}$ is retained, where the $\chi^2_{l^+l^-\mu^+\mu^-}= \chi^2_{4\rm{C}}+\Sigma^4_{i=1}\chi^2_{\rm{PID}}(i)(i = e,\mu$, or $\pi$) is the sum of the $\chi^2$ values from the 4C kinematic fit and PID.
In addition, a 4C kinematic fit under the hypothesis of $\gamma\pi^+\pi^-\pi^+\pi^-$ is also performed, and $\chi^2_{\pi^+\pi^-\pi^+\pi^-}>\chi^2_{l^+l^-\mu^+\mu^-}$ is required to reject the possible background of $J/\psi\rightarrow\gamma\pi^+\pi^-\pi^+\pi^-$.

\section{ANALYSIS OF $\eta/\eta^\prime \to e^+e^-\mu^+\mu^-$}
\label{sec:etap2e2mu}
\begin{figure}[h]
\begin{center}
\begin{minipage}[t]{0.9\linewidth}
\includegraphics[width=1\textwidth]{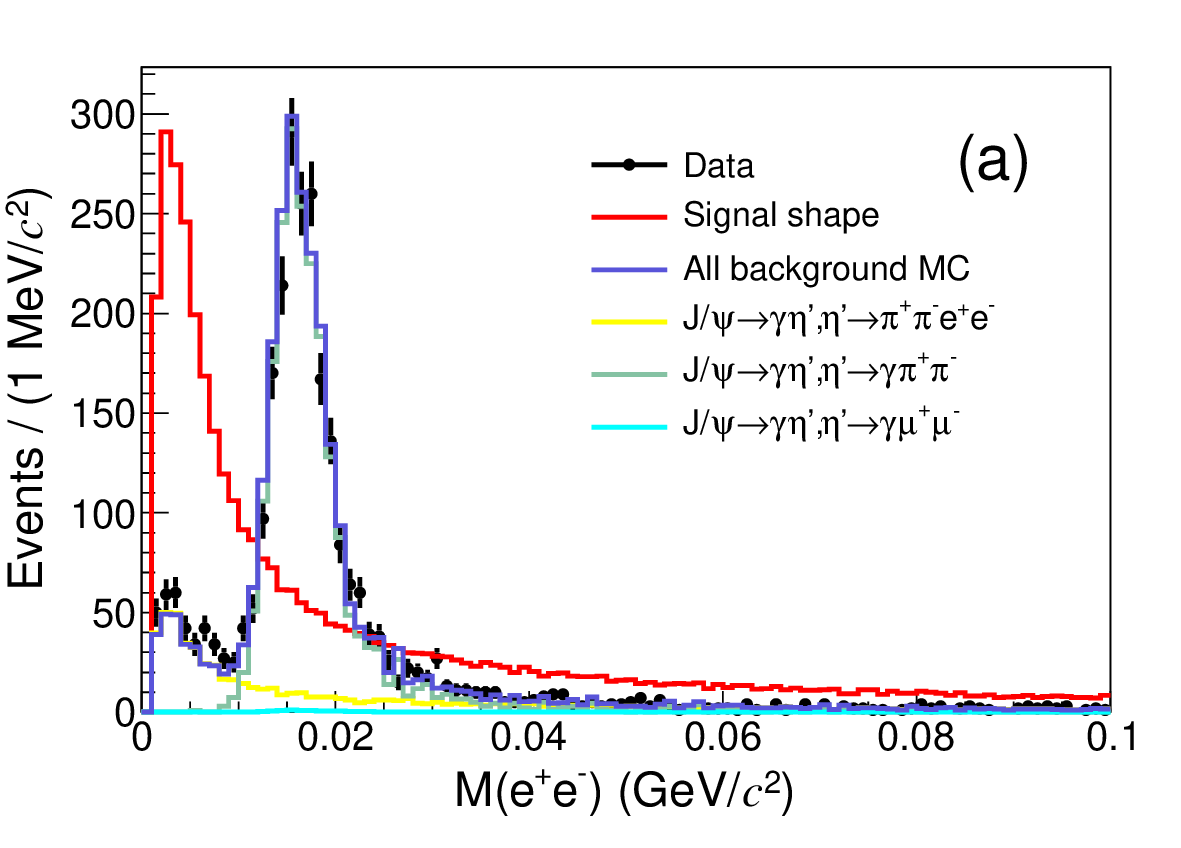}
\end{minipage}
\begin{minipage}[t]{0.9\linewidth}
\includegraphics[width=1\textwidth]{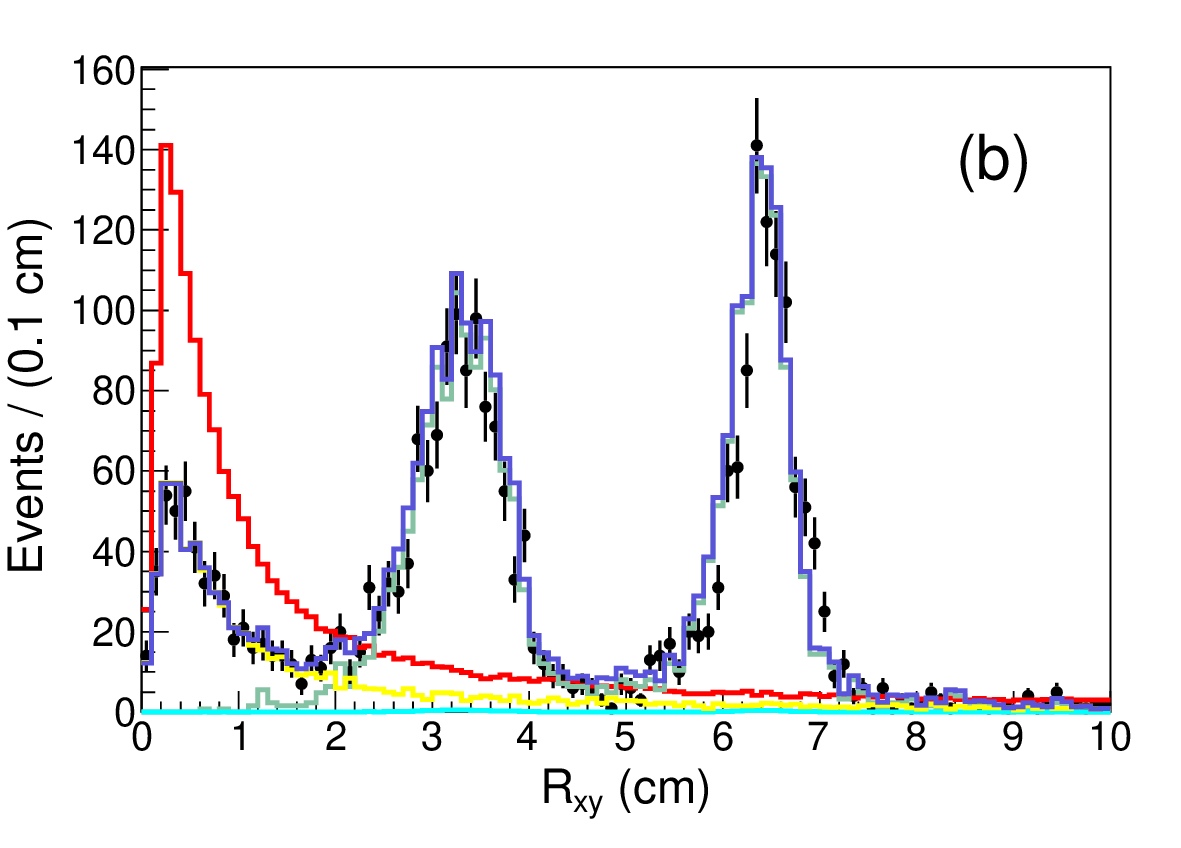}
\end{minipage}
\caption{The distributions of (a) $M_{ee}$ and (b) $R_{xy}$. The dots with error bars are data, the red histograms are the signal MC events, the yellow histograms are the background from   $\eta'\to \pi^{+}\pi^{-}e^{+}e^{-}$ and the green histograms are the background from  $\eta'\to \gamma \pi^{+}\pi^{-}$, while the blue histograms are the sum of all backgrounds estimated from the MC simulation.  The MC simulations are normalized to the yields found in Table \ref{tab_etap2e2mu_bkgs}.}
\label{fig_beforecutgconv}
\end{center}
\end{figure}

For the decay $\eta/\eta^\prime \to e^+e^-\mu^+\mu^-$, after requiring $\chi^2_{\gamma e^+e^-\mu^+\mu^-}<40$, the dominant remaining background events come from the decay $J/\psi\to\gamma\eta/\eta^\prime$ with $\eta/\eta^\prime\to\gamma\mu^+\mu^-$ and $\gamma\pi^+\pi^-$, where the photon converts to an $e^+e^-$ pair at the beam pipe or the inner wall of the MDC.  However, the BESIII tracking algorithm uses the IP as the reference point for all tracks, which leads to the directions of the tracks not originating from the IP to be shifted. As a result, the $e^+e^-$ pairs from $\gamma$ conversions have larger $M(e^+e^-)$ values than the true values. For this analysis, the $\gamma-$conversion related backgrounds appear as a large peak around 0.015 GeV/$c^2$ as shown in Fig.~\ref{fig_beforecutgconv}(a). By finding the intersection of the $e^+$- and $e^-$-helices in the $r-\phi$ plane, we can get the distance from the $e^+e^-$- vertex position to the IP, $R_{xy}$. Figure~\ref{fig_beforecutgconv}(b) shows the $R_{xy}$ distribution for the selected events, which shows three characteristic locations in the detector. The signal pairs effectively originate from the IP, and the peak around 3~cm comes from the conversion backgrounds at the beam pipe, while the peak around 6~cm comes from conversion events at the inner wall of the MDC.

\begin{figure*}[htbp]
\begin{center}
\begin{minipage}[t]{0.245\linewidth}
\includegraphics[width=1\textwidth]{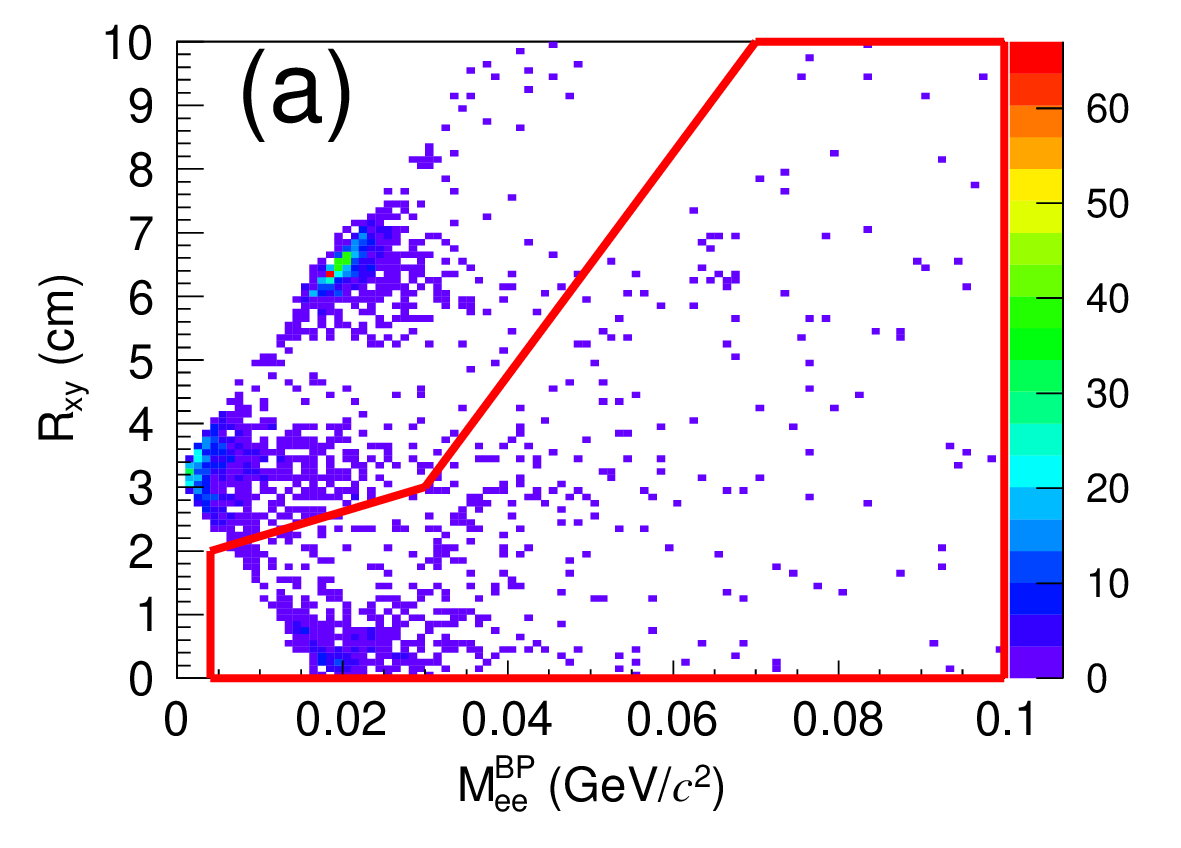}
\end{minipage}
\begin{minipage}[t]{0.245\linewidth}
\includegraphics[width=1\textwidth]{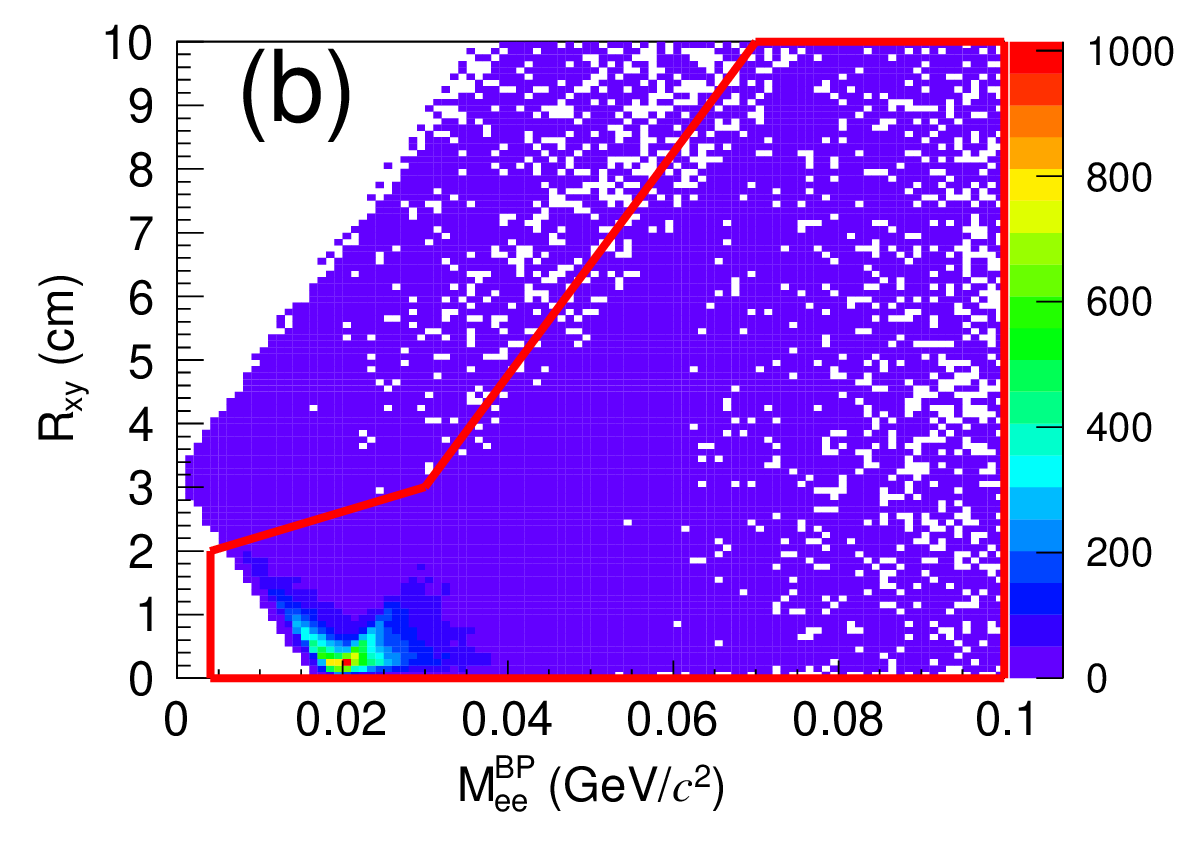}
\end{minipage}
\begin{minipage}[t]{0.245\linewidth}
\includegraphics[width=1\textwidth]{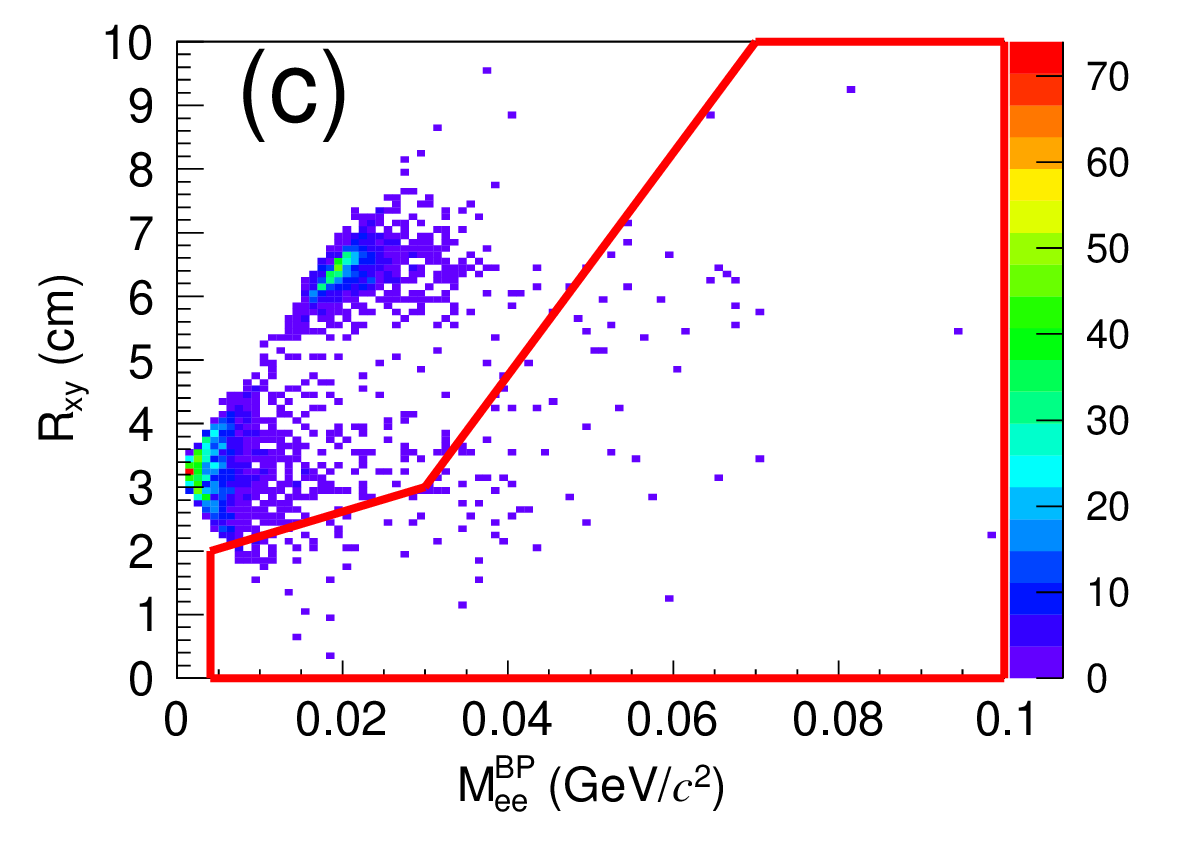}
\end{minipage}
\begin{minipage}[t]{0.245\linewidth}
\includegraphics[width=1\textwidth]{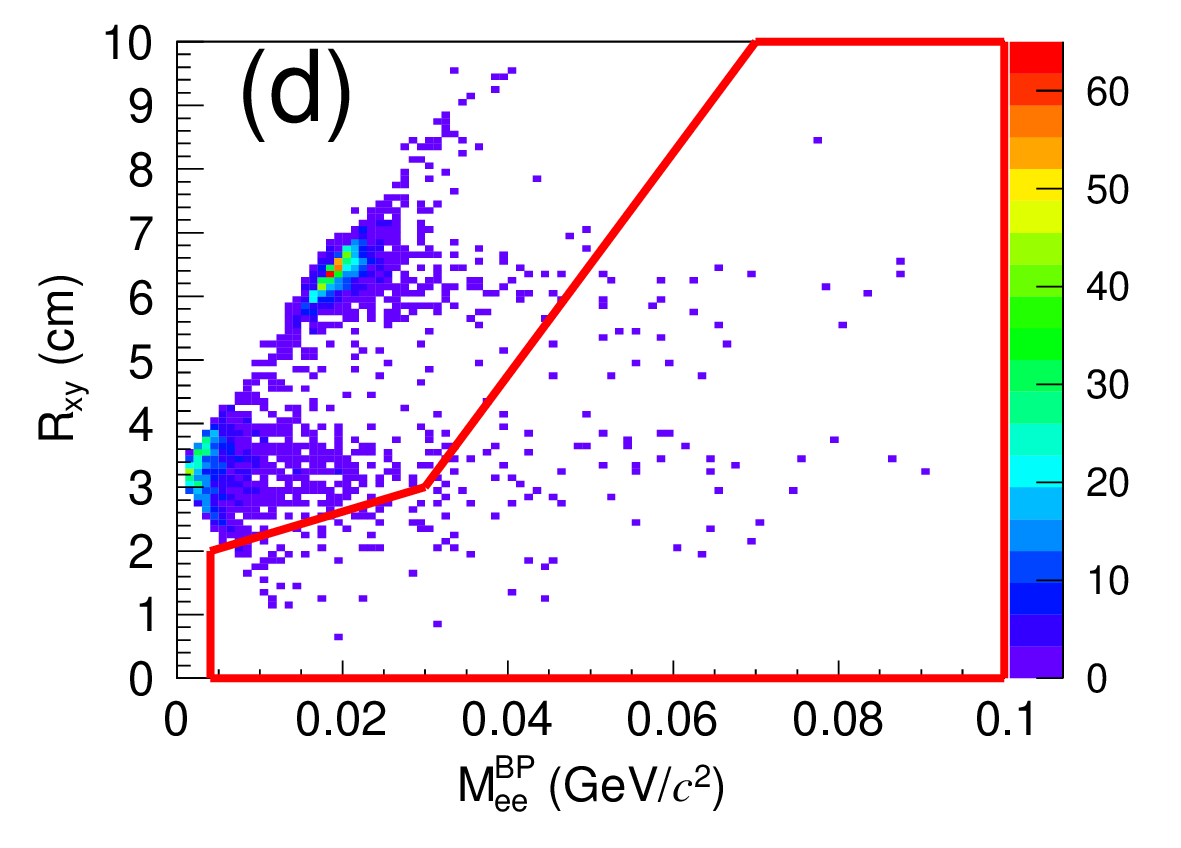}
\end{minipage}
\begin{minipage}[t]{0.245\linewidth}
\includegraphics[width=1\textwidth]{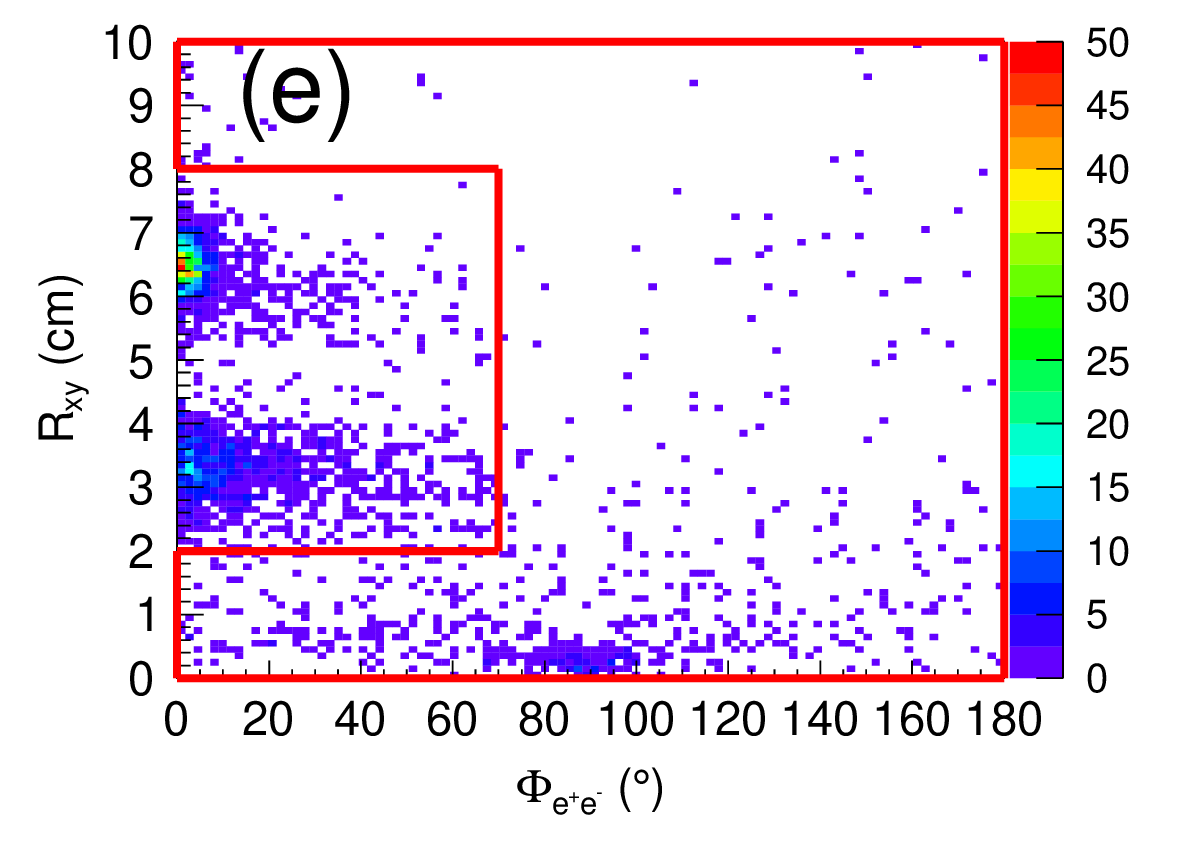}
\end{minipage}
\begin{minipage}[t]{0.245\linewidth}
\includegraphics[width=1\textwidth]{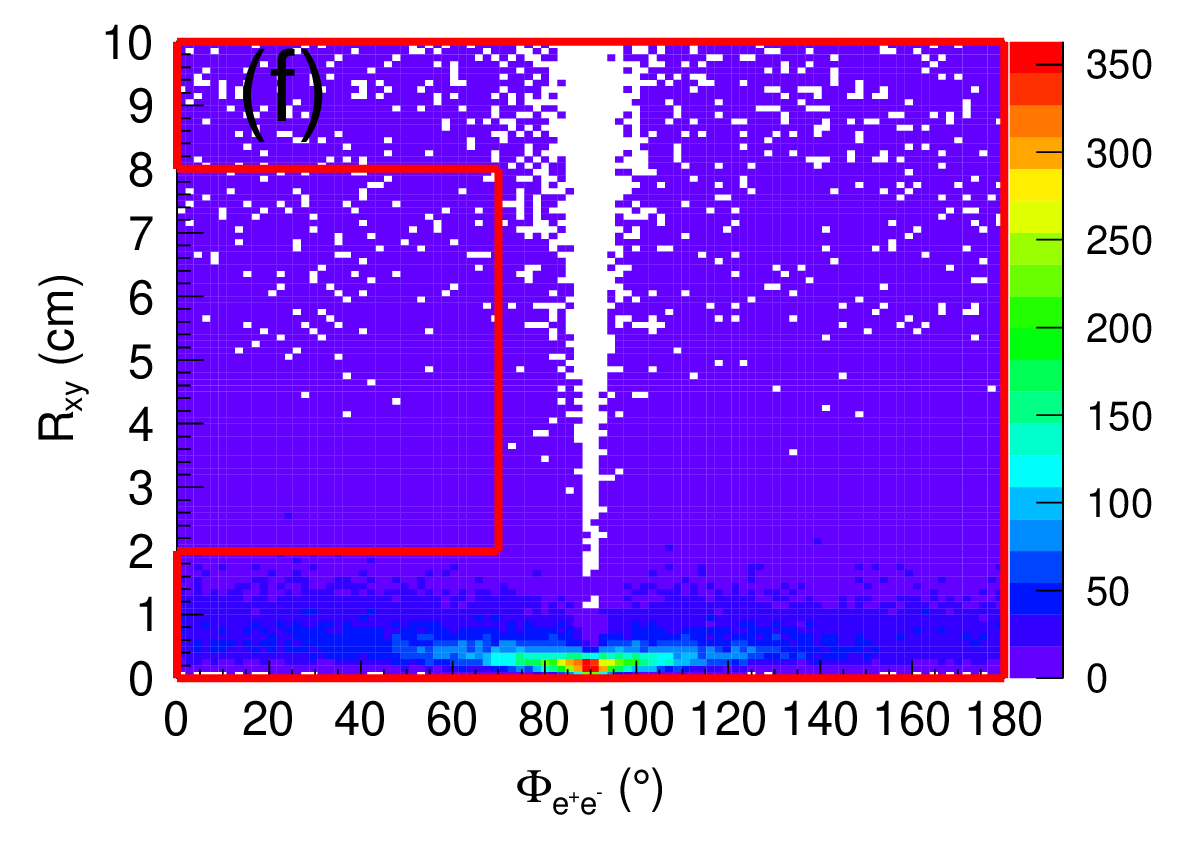}
\end{minipage}
\begin{minipage}[t]{0.245\linewidth}
\includegraphics[width=1\textwidth]{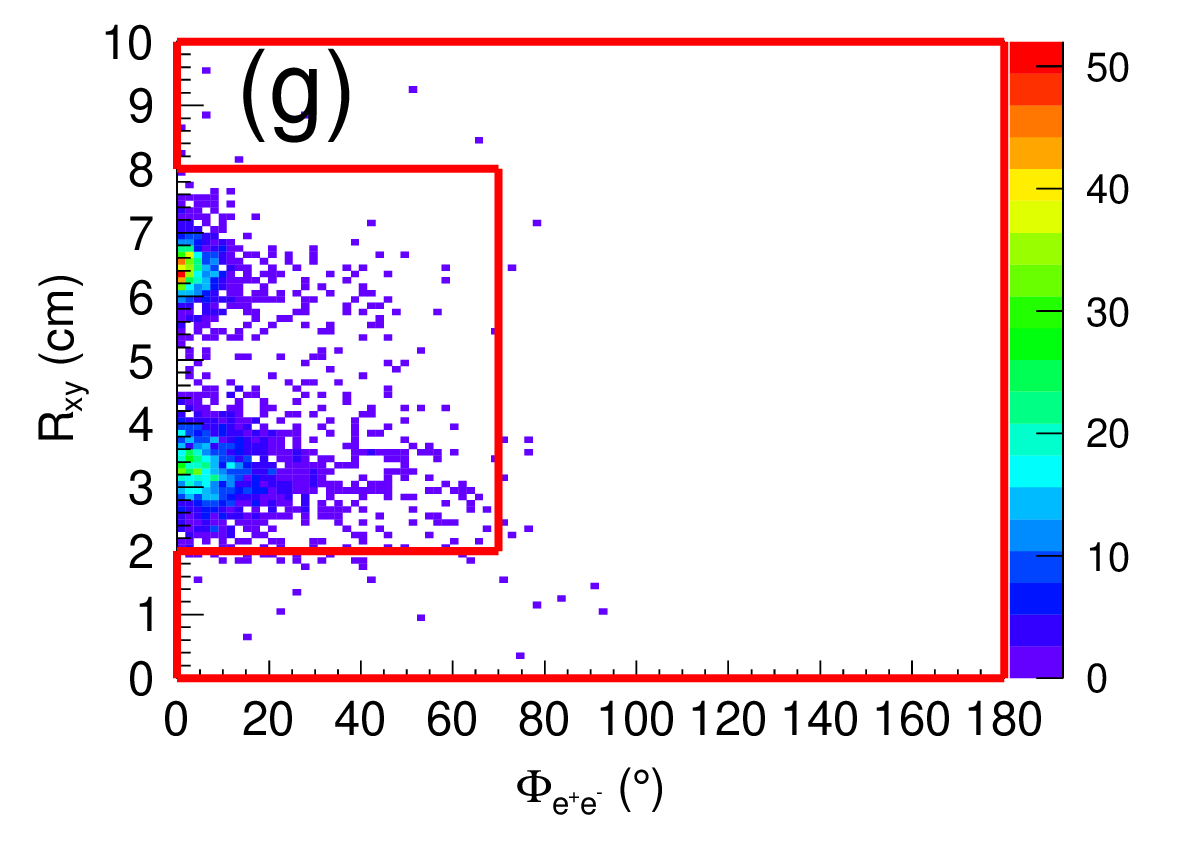}
\end{minipage}
\begin{minipage}[t]{0.245\linewidth}
\includegraphics[width=1\textwidth]{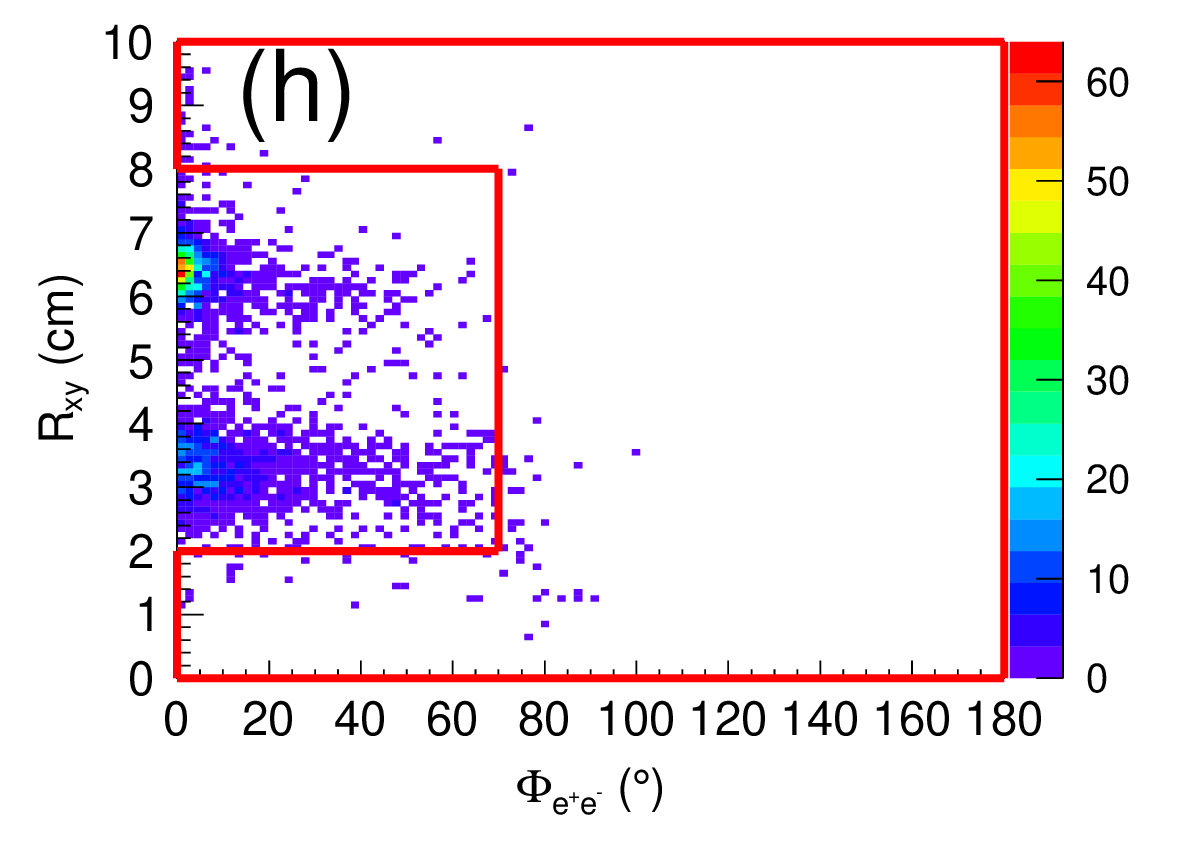}
\end{minipage}
\caption{
The distributions of $M_{ee}^{BP}$ versus $R_{xy}$~((a),(b),(c),(d)) and $\Phi_{ee}$ versus $R_{xy}$~((e),(f),(g),(h)) for data, signal MC sample, $\eta'\to \gamma \mu^{+}\mu^{-}$ MC sample and $\eta'\to \gamma \pi^{+}\pi^{-}$ MC smaple, respectively. All events outside the solid (red) polygons are rejected.
}
\label{fig_rxy_Meebp_Phiee}
\end{center}
\end{figure*}

In order to suppress photon conversion events and improve the signal-to-background ratio, two additional discriminating variables are introduced.
The first one is the invariant mass of each $e^+e^-$ pair at the beam pipe, $M_{ee}^{BP}$, which is calculated using the $e^{\pm}$ momentum vectors determined at their points of intersection with the beam pipe. This corrects the directions of the vectors but not their magnitudes. The second variable is the $z$-projection of the opening angle of the $e^+e^-$ pairs, where $z$ is the magnetic field direction, $\Phi_{ee}$. For $e^+e^-$ pairs originating from the IP, the opening angle is increased and consequently their invariant masses become larger than the true value. The momenta of the $e^+e^-$ pairs that are reconstructed in the beam pipe are approximately parallel and the invariant mass close to the minimum value, $2m_e$. In conversion events, $\Phi_{ee}$ is expected to be close to zero, whereas in the $\eta' \to e^+e^-\mu^+\mu^-$ decay it varies widely. As shown in the two-dimensional distributions of $M_{ee}^{BP}$ versus $R_{xy}$, and $\Phi_{ee}$ versus $R_{xy}$ in Fig.~\ref{fig_rxy_Meebp_Phiee}, the signal events and photon conversion events are well separated.

Photon conversion events are effectively vetoed by rejecting events with $\Phi_{ee} < 70^{\circ}$ and 2 cm$< R_{xy} <$8 cm. According to the distribution of $M_{ee}^{BP}$ versus $R_{xy}$, we select all events to the high mass side of a curve defined by straight line segments between the points (0.004 GeV/$c^2$, 0 cm), (0.004 GeV/$c^2$, 2 cm), (0.03 GeV/$c^2$, 3 cm) and (0.07 GeV/$c^2$, 10 cm).

\begin{figure*}[hptb]
\begin{center}
\begin{minipage}[t]{0.47\linewidth}
\includegraphics[width=1\textwidth]{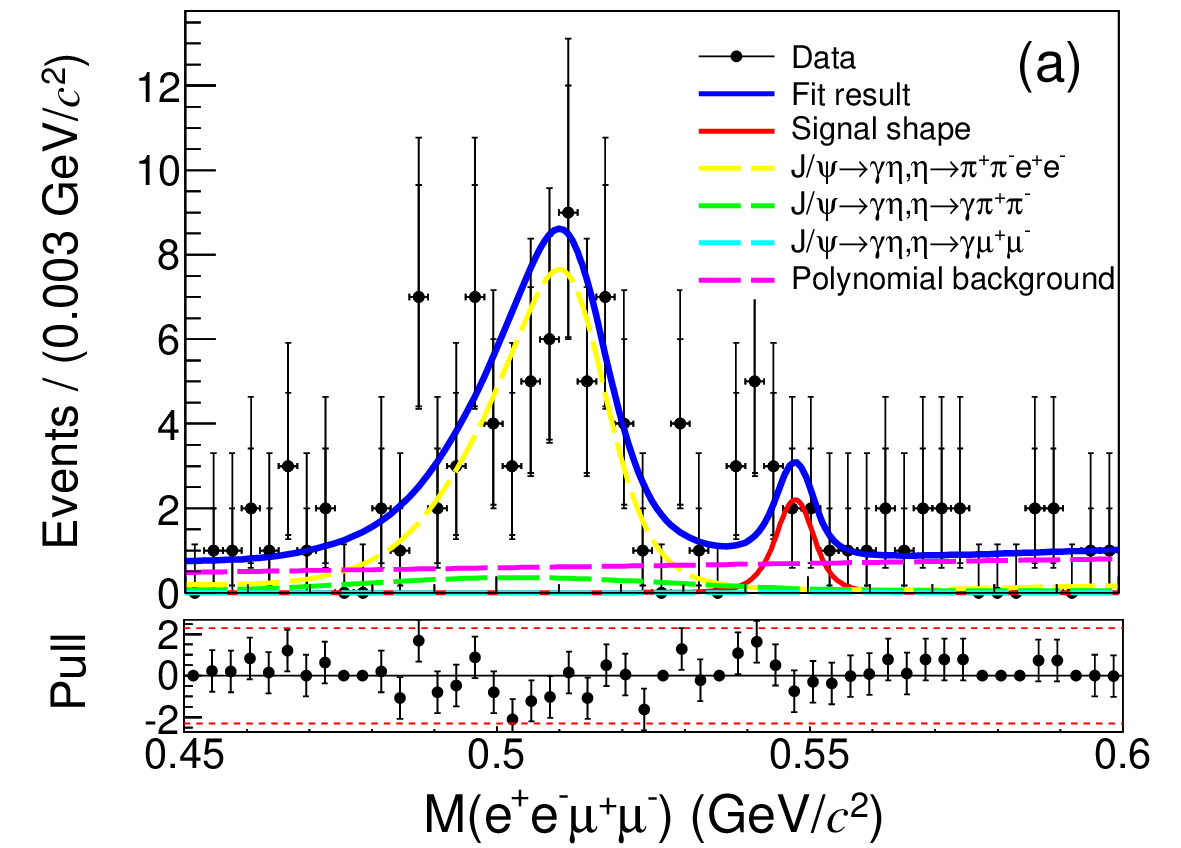}
\end{minipage}
\hspace{0.615cm}
\begin{minipage}[t]{0.47\linewidth}
\includegraphics[width=1\textwidth]
{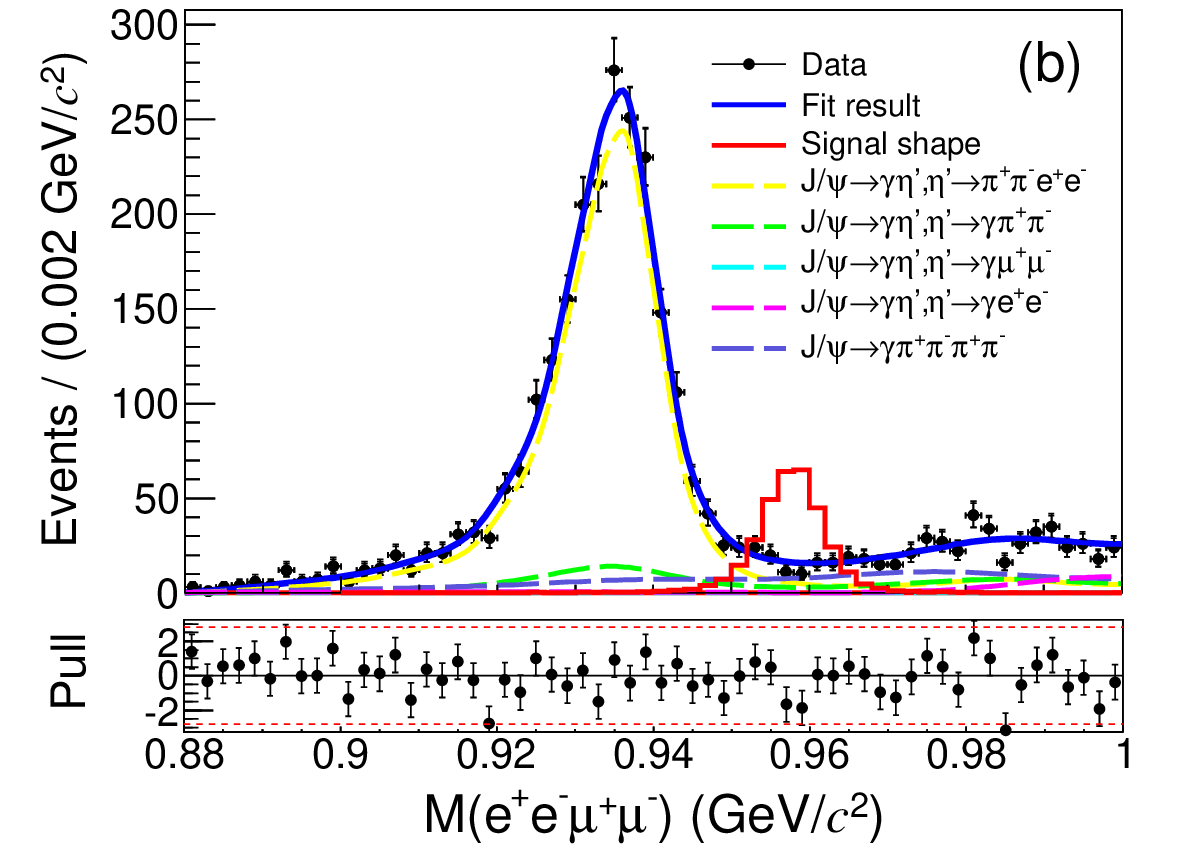}
\end{minipage}
\caption{The fits to the distributions of $M(e^{+}e^{-}\mu^{+}\mu^{-})$ for (a)~$\eta \to  e^{+} e^{-} \mu^{+} \mu^{-}$ and (b)~$\eta' \to  e^{+} e^{-} \mu^{+} \mu^{-}$. The dots with error bars are data, the red solid lines are the signal shapes, the blue curves are the fit results, and the other lines in various colors are from different backgrounds as shown on the figure. }
\label{fig4_2}
\end{center}
\end{figure*}

Figures~\ref{fig4_2}(a) and \ref{fig4_2}(b) show the $M(e^+e^-\mu^+\mu^-)$ distributions in the $\eta$ and $\eta^\prime$ mass regions, respectively, for candidates surviving all the above selection criteria.  No signals are evident for either the $\eta$ or $\eta^\prime$. 

To investigate background contributions in the $M(e^+e^-\mu^+\mu^-)$ distribution, an inclusive MC sample of 10 billion $J/\psi$ events is analyzed with the same procedure. The background events are mainly from $\eta/\eta^\prime \rightarrow \pi^+\pi^-e^+e^-, \gamma\pi^+\pi^-$ and $\gamma\mu^+\mu^-$, as summarized in Tables~\ref{tab_etap2e2mu_bkgs} and \ref{tab_eta2e2mu_bkgs}. Dedicated MC samples for these background channels are produced. Their contributions are normalized according to the branching fractions in the PDG~\cite{PDG} and fixed in the fits to Figs.~\ref{fig4_2}(a) and \ref{fig4_2}(b).

\begin{table}[h]
  \caption{The dominant background channels and the normalized yields for $\eta' \to e^+e^-\mu^+\mu^-$. A dash indicates that the corresponding contribution is free in the fit. The uncertainties are statistical.}
  \begin{tabular} {c c}
  \hline\hline
  Background channel  & Normalized yield\\
  \hline
${J/{\psi}\to \gamma\eta',  \eta'\to \pi^{+}\pi^{-}e^{+}e^{-}}$  &${3053 \pm 132}$\\
${J/{\psi}\to \gamma\eta' , \eta'\to \gamma \pi^{+}\pi^{-}}$     &${296.8\pm 5.6}$ \\
${J/{\psi}\to \gamma\eta' , \eta'\to \gamma \mu^{+}\mu^{-}}$    &${1.7  \pm 0.4}$ \\
${J/{\psi}\to \gamma\eta' , \eta'\to \gamma e^{+}e^{-}}$   &${0.8 \pm 0.1}$ \\
${J/{\psi}\to \gamma \pi^{+}\pi^{-}\pi^{+}\pi^{-}}$   & - \\

   \hline
   \hline
  \end{tabular}
  \label{tab_etap2e2mu_bkgs}
\end{table}

\begin{table}[h]
  \caption{The dominant background channels and the normalized yields for $\eta \to e^+e^-\mu^+\mu^-$. The uncertainties are statistical.}
  \begin{tabular} {c c}
  \hline\hline
  Background channel  & Normalized yield\\
  \hline
${J/{\psi}\to \gamma\eta,  \eta \to \pi^{+}\pi^{-}e^{+}e^{-}}$  &{${80.1 \pm 3.5}$}  \\
${J/{\psi}\to \gamma\eta , \eta \to \gamma \pi^{+}\pi^{-}}$    &{${9.8 \pm 0.2}$}   \\
${J/{\psi}\to \gamma\eta , \eta \to \gamma \mu^{+}\mu^{-}}$    &{${0.1 \pm 0.02}$}    \\
   \hline
   \hline
  \end{tabular}
  \label{tab_eta2e2mu_bkgs}
\end{table}

To check the statistical significances of the $\eta$/$\eta^\prime$ signals, the unbinned extended maximum likelihood fits are performed to the $M(e^+e^-\mu^+\mu^-)$ distribution. In the fit, the total probability density function
consists of a signal and various background contributions. The signal component is modeled with the simulated shape.
The background components considered are
subdivided into two classes: (i) the shapes of those
background events that contribute to a structure in the 
$M(e^+e^-\mu^+\mu^-)$ distribution around the $\eta$ (e.g., $\eta \to \pi^{+}\pi^{-}e^{+}e^{-}$, $\eta \to \gamma \pi^{+}\pi^{-}$ and $\eta \to \gamma \mu^{+}\mu^{-}$) or $\eta^\prime$ mass regions (e.g., $\eta'\to \pi^{+}\pi^{-}e^{+}e^{-}, \eta'\to \gamma \pi^{+}\pi^{-}, \eta'\to \gamma \mu^{+}\mu^{-}, \eta'\to \gamma e^{+}e^{-}$) are derived from the dedicated MC
samples; (ii) the combinatorial background shape of $J/\psi\rightarrow\gamma\pi^+\pi^-\pi^+\pi^-$ is also described with the simulated shape, and its magnitude is left free in the fit.

Based on the differences of the likelihood values obtained by including or not including the signal component in the fits, and considering the changes in the degrees of freedom, the statistical significances for both the $\eta$ and $\eta^\prime$ signals are estimated to be less than $3\sigma$. 

We then use the Bayesian approach~\cite{Bayesian} to determine the upper limits on the $\eta$/$\eta^\prime$ signal yields. A series of unbinned extended maximum likelihood fits are performed to the $M(e^+e^-\mu^+\mu^-)$ distribution with an expected $\eta$/$\eta^\prime$ signal yield. The distribution of normalized likelihood values, defined as $\mathcal{L}(N)={\rm{exp}}(-[\mathcal{S}(N) - {\mathcal{S}}_{\rm{{min}}}])$, where ${\mathcal{S}}_{\rm{{min}}}$ is the lowest negative log-likelihood obtained from the ensemble of fits, is directly taken as the probability density function.
The upper limit on the number of signal events at the 90\% confidence level~(C.L.) is defined as $N^{\rm{UL}}$
corresponding to the number of signal events at 90\% of the
integral of the probability density function,
\begin{equation}
\label{eqBr0_BF2}
 {\textstyle{{\int}^{{N}^{\rm{UL}}}_0 {\mathcal{L}}(N) d{N} \over {\int}^{\infty}_0 {\mathcal{L}}(N) d {N}}} = 0.9,
\end{equation}
where $N$ is the expected number of signal events.  
To take into account the additive uncertainties associated with the fit to data, we perform a series of alternative fits by varying the fit ranges, continuum background shapes and peaking background shapes to extract the corresponding upper limits of the signal yields. For the $\eta$ and $\eta^\prime$ cases, the maximum upper limits are 15.9 and 19.0, respectively.

\section{ANALYSIS OF $\eta/\eta' \to \mu^+\mu^-\mu^+\mu^-$}
\label{sec:etap4mu}
\begin{figure}[h]
\begin{center}
\begin{minipage}[t]{0.97\linewidth}
\includegraphics[width=1\textwidth]{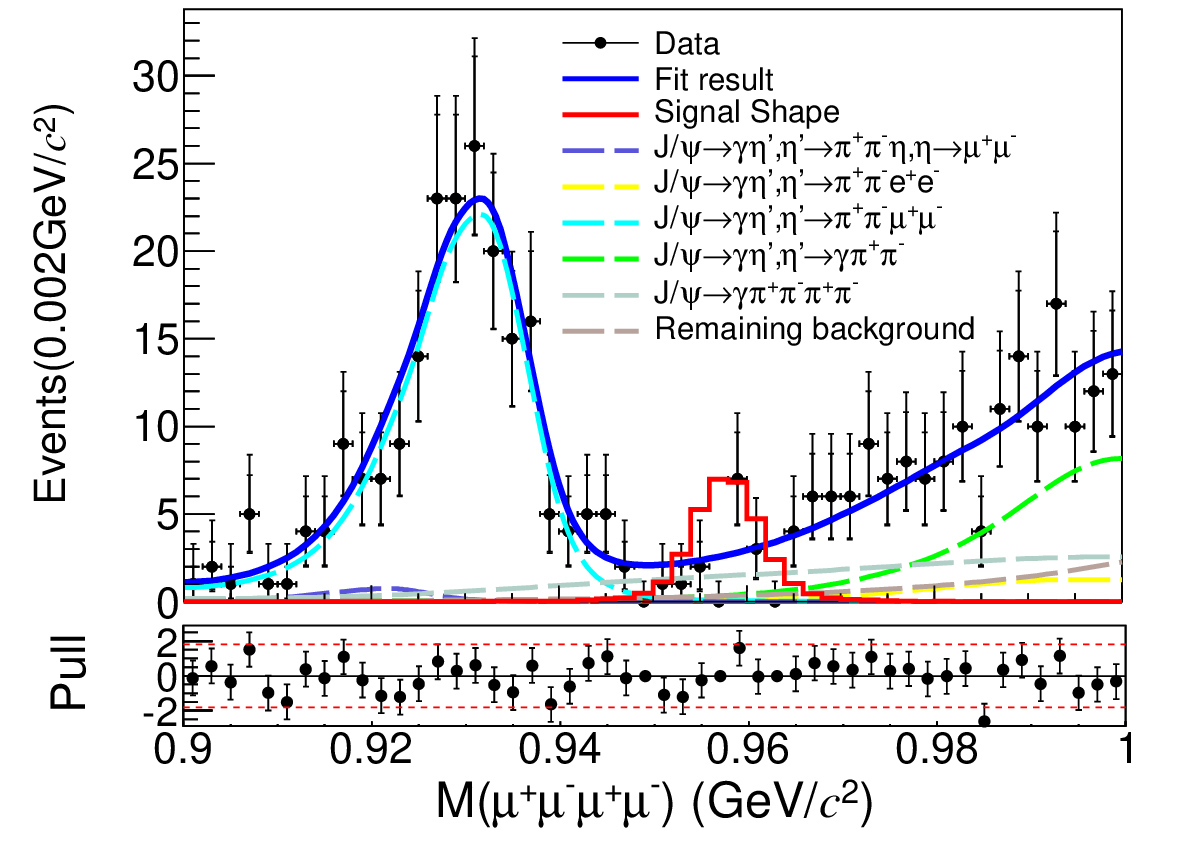}
\end{minipage}
\caption{The fit to the distribution of $M(\mu^{+}\mu^{-}\mu^{+}\mu^{-})$. The dots with error bars are data, the red solid line is the signal shape, the blue curve is the fit result, and the other lines in various colors are from different backgrounds as shown on the figure.}
\label{fig4_3}
\end{center}
\end{figure}

For the decay $\eta' \to \mu^+\mu^-\mu^+\mu^-$, after requiring $\chi^2_{\gamma \mu^+\mu^-\mu^+\mu^-}<40$, the distribution of $M(\mu^+\mu^-\mu^+\mu^-)$ is shown in Fig.~\ref{fig4_3}. 
The background events come mainly from $J/{\psi}\to \gamma\eta' ( \eta'\to \pi^{+}\pi^{-}\mu^{+}\mu^{-},  \pi^{+}\pi^{-}e^{+}e^{-}, \gamma \pi^{+}\pi^{-} $ and $\pi^{+} \pi^{-} \eta,  \eta\rightarrow\mu^+\mu^-)$. The yields of each background channel are estimated with the dedicated MC samples according to the corresponding branching fractions in the PDG~\cite{PDG}, as summarized in Table~\ref{tab_etap4mu_bkgs}. The remaining background is represented by an exponential function.

The upper limit on the signal yield is obtained from a series of unbinned extended maximum likelihood fits to the $M(\mu^{+}\mu^{-}\mu^{+}\mu^{-})$ spectrum. In the fit, as shown in Fig.~\ref{fig4_3}, the line shape of the $\eta^\prime$ signal is determined by MC simulation, the shapes of the dominant background contributions listed in Table~\ref{tab_etap4mu_bkgs} are also from the MC simulations, while the remaining background contribution is described with the phase space modeled MC sample of $J/\psi\rightarrow\gamma \pi^+\pi^-\pi^+\pi^-$. The upper limit of the signal yield at the 90\% C.L. is set to be 11.7 with the same method as described in section IV. 

There is no event in the $\eta$ mass region for $\eta \to \mu^+\mu^-\mu^+\mu^-$ after the event selection, an upper limit on the branching fraction is estimated at approximately $10^{-7}$.

\begin{table}[h]
  \caption{The dominant background channels and the normalized yields for $\eta' \to \mu^+\mu^-\mu^+\mu^-$. A dash indicates that the number of background events is free in the fit. The uncertainties are statistical.}
  \begin{tabular} {c c}
  \hline\hline
  Background channel  & Normalized yield\\
  \hline
${J/{\psi}\to \gamma\eta' , \eta'\to \pi^{+}\pi^{-}\mu^{+}\mu^{-}}$  &${220.6 \pm 44.2}$\\
${J/{\psi}\to \gamma\eta' , \eta'\to \pi^{+}\pi^{-}e^{+}e^{-}}$     &{${30.7 \pm 1.3}$} \\
${J/{\psi}\to \gamma\eta' , \eta'\to \gamma \pi^{+}\pi^{-}}$    &{${41.7\pm 0.8}$} \\
$J/{\psi} \to \gamma \eta', \eta' \to \pi^{+} \pi^{-} \eta, \eta \to \mu^{+}\mu^{-}$    &{${11.3 \pm 1.6}$}\\
${J/{\psi}\to \gamma \pi^{+}\pi^{-}\pi^{+}\pi^{-}}$   & - \\

   \hline
   \hline
  \end{tabular}
  \label{tab_etap4mu_bkgs}
\end{table}

\section{SYSTEMATIC UNCERTAINTY}
\label{sec:systematics}
The systematic uncertainties of the branching fractions are discussed below. They are assumed to be independent and are summed in quadrature to extract the total uncertainties. The corresponding contributions as summarized in Table \ref{list_sys} are discussed in detail below.
\begin{itemize}
\item Number of $J/{\psi}$ events:
The total number of $J/{\psi}$ events in data is determined to be $(10087 \pm 44) \times {10^{6}}$ by counting the inclusive hadron events, and its uncertainty is 0.4\% ~\cite{EVENTS}.

\item Branching fractions of $J/{\psi} \to \gamma \eta$ and $J/{\psi} \to \gamma \eta'$:
The uncertainties, $1.7\%$ and $1.3\%$, are taken from the PDG~\cite{PDG}.

\item MDC tracking:
The uncertainty is determined to be 1.0\% per track for electrons, using a control sample of $ e^+e^- \to e^+e^-\gamma$. Since there is no appropriate control sample to study the tracking for muons in the low momentum region, we also take 1.0\% per track as the tracking uncertainty for muons. The total systematic uncertainty due to the MDC tracking efficiency is assigned to be 4.0\% for the four charged tracks in the decay $\eta$/$\eta'\to l^+l^-\mu^+\mu^-$.

\item PID:
The systematic uncertainty in the PID efficiency is evaluated in the same way as that for the tracking efficiency, and is also assigned to be 4.0\% for each signal decay.

\item Photon detection:
The systematic uncertainty from the photon reconstruction has been studied extensively using the process $e^+e^- \to \gamma_{\rm{ISR}} \mu^+\mu^-$, where ISR stands for initial state radiation. The difference in efficiency between data and MC simulation, up to 0.5\%, is assigned as the systematic uncertainty.

\item Kinematic fit:
To investigate the systematic uncertainty associated with the kinematic fit, the track helix parameters correction method~\cite{4C} is used. Half of the difference in the detection efficiency with and without this correction is taken as the systematic uncertainty.

\item Photon conversion veto:
The systematic uncertainty from the photon conversion veto has been studied with a clean control sample of $J/\psi\to\pi^+\pi^-\pi^0$, $\pi^0\to\gamma e^+e^-$~\cite{DIY2}. The relative differences of efficiencies associated with the photon conversion rejection between data and MC simulation are assigned as the systematic uncertainties, which are 1.2\% for $\eta' \to e^+e^-\mu^+\mu^-$ and 0.8\% for $\eta \to e^+e^-\mu^+\mu^-$.

\item Generator model:
The MC generator based on the theoretical calculation as explained in Ref.~\cite{LUN} is used to determine the detection efficiency. The detection efficiency dependence on the form factor is evaluated by replacing the form factor above with the form factors introduced in the modified VMD model described in Ref.~\cite{VMD}. The maximum difference of the detection efficiency between the nominal and alternative models is taken as the uncertainty.
\end{itemize}
\begin {table}[h]
\begin{center}
\begin{small}
\caption {Relative systematic uncertainties (in \%). The symbols I, II and III represent the decay modes of $\eta' \to  e^{+} e^{-} \mu^{+} \mu^{-}$,  $\eta \to  e^{+} e^{-} \mu^{+} \mu^{-}$  and  $\eta' \to  \mu^{+} \mu^{-}\mu^{+} \mu^{-}$, respectively. }
\label{list_sys}
\begin {tabular}{l c c c c}\hline\hline

Source & ~~~I~~~ & ~~~II~~~ & ~~~III~~~      \\
 \hline
Number of $J/{\psi}$ events &  $0.4$       & $0.4$   &  $0.4$                           \\
$\mathcal {B}(J/{\psi}\to\gamma \eta / \eta')$	&  $1.3$     &  $1.7$   &  $1.3$           \\
MDC tracking  &     $4.0$      &  $4.0$     &     $4.0$     \\
PID           &     $4.0$      &  $4.0$     &     $4.0$     \\
Photon detection    &     $0.5$      &  $0.5$   &    $0.5$ \\
Kinematic fit                   &     $0.5$     &  $0.4$    &     $0.4$ \\
Photon conversion veto    &  $1.2$       &  $0.8$      &     $-$  \\
Generator model           &  $0.2$       & $0.5$       &     $0.1$\\
\hline
	Total                 & $6.0$        &  $6.1$    & $5.9$ \\
\hline
\hline
\end{tabular}
\end{small}
\end{center}
\end{table}
\section{RESULTS}
\begin{figure*}[htbp]
\begin{center}
\begin{minipage}[t]{0.325\linewidth}
\includegraphics[width=1\textwidth]{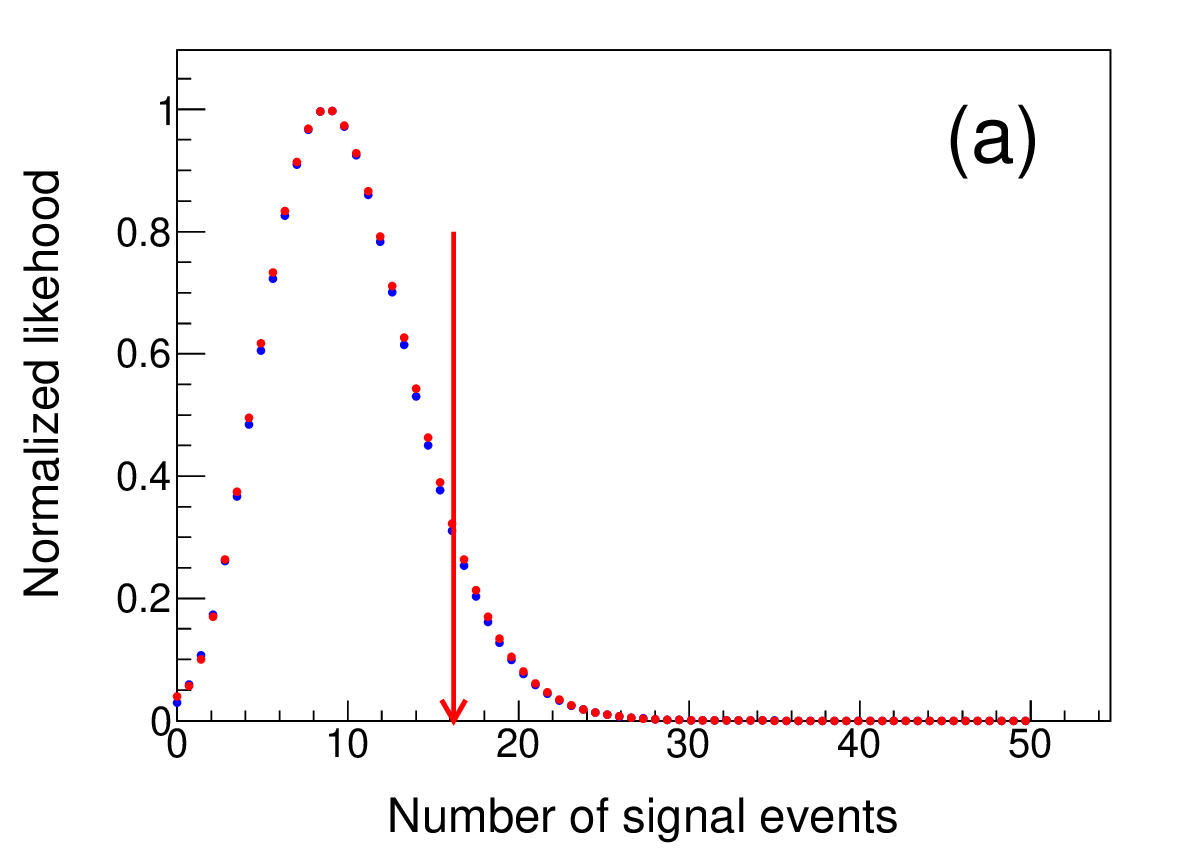}
\end{minipage}
\begin{minipage}[t]{0.325\linewidth}
\includegraphics[width=1\textwidth]{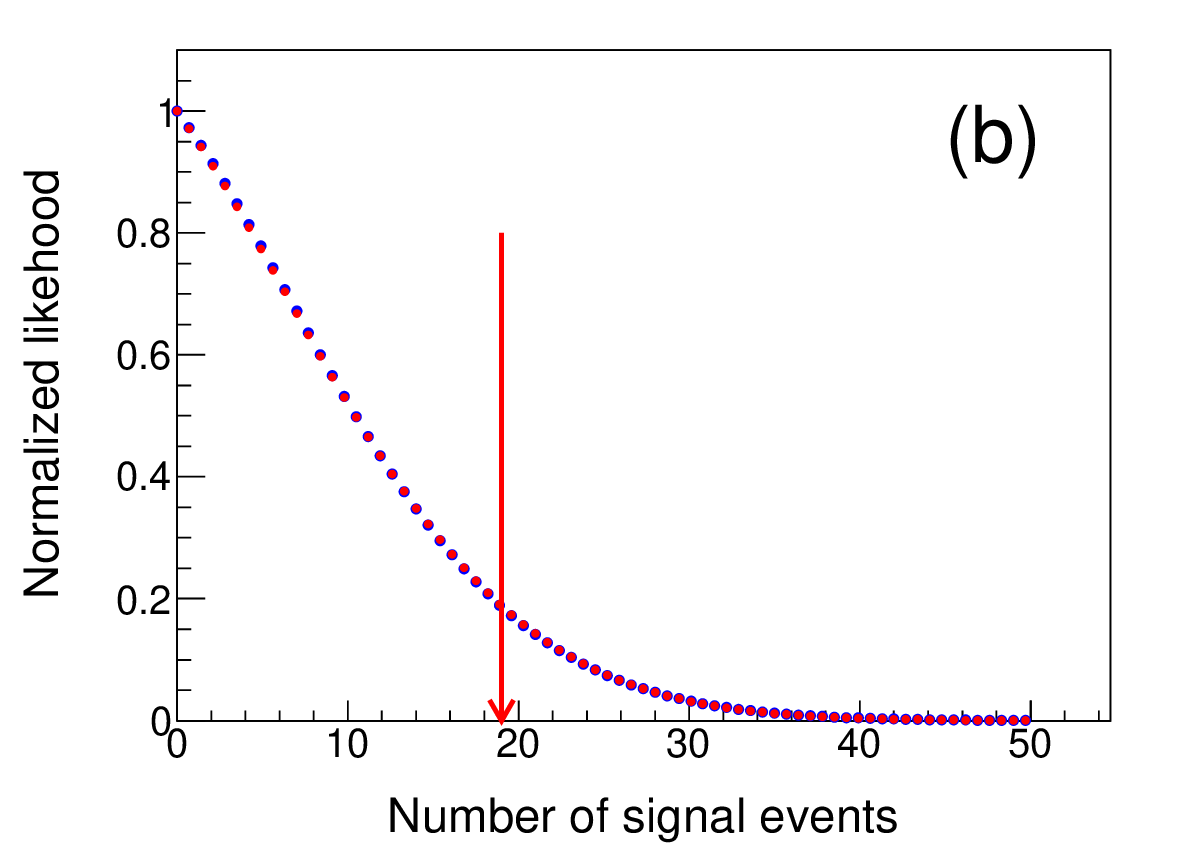}
\end{minipage}
\begin{minipage}[t]{0.325\linewidth}
\includegraphics[width=1\textwidth]{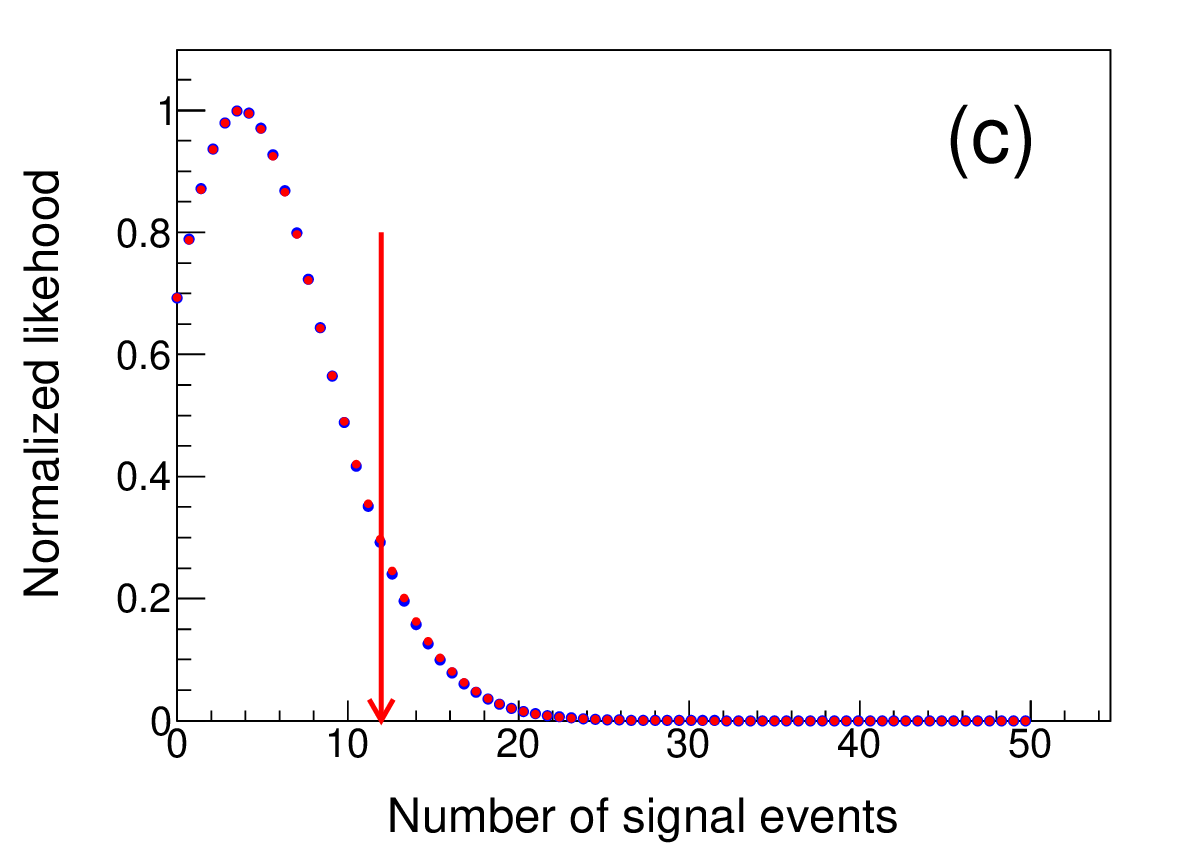}
\end{minipage}
\caption{Normalized likelihood distributions for (a)~$\eta\rightarrow e^+e^-\mu^+\mu^-$, (b)~$\eta^\prime\rightarrow e^+e^-\mu^+\mu^-$ and (c)~$\eta^\prime\rightarrow \mu^+\mu^-\mu^+\mu^-$. The blue and red dots denote those without and with incorporating systematic uncertainties.}
\label{smear}
\end{center}
\end{figure*}
The final upper limit on the branching fraction is determined by convolving the likelihood distribution with the multiplicative systematic uncertainty to obtain the smeared likelihood $L^\prime$,
\begin{equation}\label{eqBr0_BF2}
      L'(N)={\textstyle{{\int}^{1}_{0} L [ ({\mathcal{S}}/{\bar{\mathcal{S}}}) N]~  {\rm exp}\left[-\frac{(\mathcal{S} - \bar{\mathcal{S}})^{2}}{2 {\sigma}^2_S}  \right]  d \mathcal{S}  }},
\end{equation}
where $\mathcal{S}$ and $\bar{\mathcal{S}}$ are the detection efficiency and the nominal efficiency, respectively; $L$ is the likelihood curve without taking into account the systematic uncertainty; and $\sigma_{S}$ is the multiplicative systematic uncertainty. 

Figure~\ref{smear} shows the normalized likelihood distributions for the signal processes after taking all systematic uncertainties into account. The corresponding upper limits of the signal yields are summarized in Table~\ref{last_00} and used to evaluate the upper limits on the branching fractions. Using the detection efficiencies~($\varepsilon$) obtained from the dedicated MC simulation, the upper limit of the branching fraction for $\eta/\eta' \to l^{+}l^{-}\mu^{+} \mu^{-}$($l= e, \mu$) is calculated using
\begin{equation}
\mathcal{B}( \eta/\eta' \to l^{+} l^{-}\mu^{+} \mu^{-}  ) < {\textstyle{N^{\rm{UL}} \over {{N_{J/{\psi}}} \cdot  {\mathcal{B}}( {{J/{\psi}} \to \gamma \eta/\gamma \eta' } ) \cdot \varepsilon }}}, 
\end{equation}
where $N^{\rm{UL}}$ is the upper limit on the number of signal events, $N_{J/{\psi}}$ is the total number of $J/\psi$ events in data, $(10087 \pm 44) \times {10^{6}}$~\cite{EVENTS}, and
$\mathcal{B}( {J/{\psi}} \to \gamma \eta/\gamma \eta')$ 
is the branching fraction of ${J/{\psi}} \to \gamma \eta/\gamma \eta'$~\cite{PDG}. The obtained upper limits on the branching fractions are:
\begin{equation}
\mathcal{B}(  \eta \to  e^{+} e^{-} \mu^{+} \mu^{-}) <  6.88 \times {10^{ - 6}},
\end{equation}
\begin{equation}
\mathcal{B}(  \eta' \to  e^{+} e^{-} \mu^{+} \mu^{-}) < 1.75 \times {10^{ - 6}},
\end{equation}
\begin{equation}
\mathcal{B}(  \eta' \to \mu^{+} \mu^{-} \mu^{+} \mu^{-}) <  5.28 \times {10^{ - 7}}.
\end{equation}

\section{SUMMARY}
\label{sec：summary}
With a data sample of $(10087 \pm 44) \times {10^{6}}$ $J/{\psi}$ events, we search for the decays $\eta$/ $\eta' \to e^{+}e^{-}\mu^{+}\mu^{-}$ and $\eta' \to \mu^{+}\mu^{-}\mu^{+}\mu^{-}$. Since no evident $\eta$/$\eta'$ signal is observed, upper limits on the branching fractions for these decays at the 90\% C.L. are set, as summarized in Table \ref{last_00}. 
\begin{table}[htbp]
\begin{small}
\begin{center}
\caption{The numerical results for ${\eta}$/${\eta'}\to {l}^{+}{l}^{-}{\mu}^{+}{\mu}^{-}$($l= e, \mu$).}
\label{last_00}
\begin{tabular}{ c c c c }
     \hline
     \hline
Decay mode&  $\varepsilon$~(\%)&  $N^{\rm{UL}}$&  $\mathcal {B}^{\rm{UL}}$($90\%$ C.L.)   \\
     \hline
${\eta\to e^{+}e^{-}\mu^{+}\mu^{-}}$        &$21.50\pm0.06$  &16.2  &$6.88\times{10^{-6}}$    \\
${\eta'\to e^{+}e^{-}\mu^{+}\mu^{-}}$      &$20.54\pm0.06$  &19.0   &$1.75\times{10^{-6}}$   \\
${\eta'\to \mu^{+}\mu^{-}\mu^{+}\mu^{-}}$ &$42.91\pm0.07$  &12.0   &$5.28\times{10^{-7}}$   \\
    \hline
    \hline
\end{tabular}
\end{center}
\end{small}
\end{table}
Compared with the previous measurement, the upper limit for $\eta \to e^{+}e^{-}\mu^{+}\mu^{-}$ is improved by two orders of magnitude, while the upper limits for  $\eta^\prime\rightarrow e^+e^-\mu^+\mu^-$ and  $\eta^\prime\rightarrow \mu^+\mu^-\mu^+\mu^-$ are experimentally set for the first time. The current limits are close to the theoretical predictions as listed in Table~\ref{list0} and more data is strongly needed to explore the TFFs and to test different theoretical models.

Designed to operate with a luminosity two orders of magnitude higher than BEPCII, the future super $\tau$-charm facility~\cite{Achasov:2023gey}, will provide an excellent opportunity 
to improve the measurements of these decays.
\begin{acknowledgments}

The BESIII Collaboration thanks the staff of BEPCII and the IHEP computing center for their strong support. This work is supported in part by National Key R\&D Program of China under Contracts Nos. 2020YFA0406300, 2020YFA0406400, 2023YFA1606000; National Natural Science Foundation of China (NSFC) under Contracts Nos. 11635010, 11735014, 11935015, 11935016, 11935018, 12025502, 12035009, 12035013, 12061131003, 12192260, 12192261, 12192262, 12192263, 12192264, 12192265, 12221005, 12225509, 12235017, 12361141819; the Chinese Academy of Sciences (CAS) Large-Scale Scientific Facility Program; the CAS Center for Excellence in Particle Physics (CCEPP); Joint Large-Scale Scientific Facility Funds of the NSFC and CAS under Contract No. U1832207; 100 Talents Program of CAS; The Institute of Nuclear and Particle Physics (INPAC) and Shanghai Key Laboratory for Particle Physics and Cosmology; German Research Foundation DFG under Contracts Nos. FOR5327, GRK 2149; Istituto Nazionale di Fisica Nucleare, Italy; Knut and Alice Wallenberg Foundation under Contracts Nos. 2021.0174, 2021.0299; Ministry of Development of Turkey under Contract No. DPT2006K-120470; National Research Foundation of Korea under Contract No. NRF-2022R1A2C1092335; National Science and Technology fund of Mongolia; National Science Research and Innovation Fund (NSRF) via the Program Management Unit for Human Resources \& Institutional Development, Research and Innovation of Thailand under Contracts Nos. B16F640076, B50G670107; Polish National Science Centre under Contract No. 2019/35/O/ST2/02907; Swedish Research Council under Contract No. 2019.04595; The Swedish Foundation for International Cooperation in Research and Higher Education under Contract No. CH2018-7756; U. S. Department of Energy under Contract No. DE-FG02-05ER41374.



\end{acknowledgments}

\bibliographystyle{apsrev4-2}
\bibliography{myref}

\end{document}